\documentclass[a4, conference]{IEEEtran}      

\IEEEoverridecommandlockouts                              

\usepackage{newtxtext}
\usepackage[italic,defaultmathsizes,symbolgreek]{mathastext}
\usepackage{amssymb, amsmath}
\usepackage{nccmath}
\usepackage{hyperref}
\usepackage{cite}
\usepackage{cleveref}

\def\real{\mathcal{R}}

\newcommand{\tp}{^\mathsf{T}}

\newcommand{\bdg}{\boldsymbol{g}}

\newcommand{\bdx}{\boldsymbol{x}}

\newcommand{\bdu}{\boldsymbol{u}}

\newcommand{\bdA}{\boldsymbol{A}}
\newcommand{\bdB}{\boldsymbol{B}}

\newcommand{\bdF}{\boldsymbol{F}}

\newcommand{\bdQ}{\boldsymbol{Q}}
\newcommand{\bdR}{\boldsymbol{R}}

\newcommand{\bdcX}{\boldsymbol{\mathcal{X}}}

\newcommand{\bdcU}{\boldsymbol{\mathcal{U}}}

\newcommand{\bdvOmega}{\boldsymbol{\varOmega}}

\newcommand{\x}{x}
\newcommand{\y}{y}
\newcommand{\z}{z}

\newcommand{\vecx}{\vec{\x}}
\newcommand{\vecy}{\vec{\y}}
\newcommand{\vecz}{\vec{\z}}

\newcommand{\localframe}{L}

\newcommand{\force}{\bdF}
\newcommand{\collectivethrust}{\force_T}

\newcommand{\torquecollectivex}{\tau_{R_x}}

\newcommand{\setx}{\bdcX}
\newcommand{\setxF}{\bdF^x}
\newcommand{\setxg}{\bdg^x}
\newcommand{\setu}{\bdcU}
\newcommand{\setuF}{\bdF^u}
\newcommand{\setug}{\bdg^u}
\usepackage[    
            protrusion=true,
            expansion=true,
            final,
            babel
                ]{microtype}

\usepackage{algorithm}
\usepackage{algorithmicx}
\usepackage{algpseudocode}

\newcommand{\orcid}[1]{\href{https://orcid.org/#1}{\includegraphics[width=8pt]{img/orcid}}}

\usepackage{graphicx}
\usepackage{subcaption}
\usepackage{caption}
\usepackage{tikz}
\usepackage{tikz-3dplot}
\usepackage{pgfplots}
\pgfplotsset{compat=1.18}
\pgfmathsetmacro{\plotwidth}{0.8}
\pgfmathsetmacro{\plotheight}{0.7}
\pgfmathsetmacro{\plotheightratio}{0.8}
\pgfmathsetmacro{\subfigratio}{0.9}
\pgfmathsetmacro{\subfigwidth}{0.4}

\definecolor{color1}{HTML}{529DCB}
\definecolor{color2}{HTML}{ECA063}
\definecolor{color3}{HTML}{71BF50}
\definecolor{color4}{HTML}{F3CC4F}
\definecolor{color5}{HTML}{D46934}
\definecolor{color6}{HTML}{A1D8B6}
\definecolor{color7}{HTML}{D2C48E}
\definecolor{color8}{HTML}{F45F40}
\definecolor{color9}{HTML}{F9AE8D}
\definecolor{color10}{HTML}{80B9CE}

\def\BibTeX{{\rm B\kern-.05em{\sc i\kern-.025em b}\kern-.08em
    T\kern-.1667em\lower.7ex\hbox{E}\kern-.125emX}}

\begin{document}

\title{Trajectory Tracking for~UAVs: An Interpolating Control Approach}


	\author{\IEEEauthorblockN{Zden\v{e}k Bou\v{c}ek, Miroslav Fl\'\i dr, and Ond\v{r}ej Straka}\\
		\IEEEauthorblockA{New Technologies for the Information Society Research Center \& Department of Cybernetics\\
			Faculty of Applied Sciences\\
			University of West Bohemia\\
			Pilsen, Czechia\\
			Email: zboucek@kky.zcu.cz, flidr@kky.zcu.cz, straka30@kky.zcu.cz}}
   
	

\maketitle

\begin{abstract}
Building on~our previous work, this paper investigates the~effectiveness of~interpolating control (IC) for~real-time trajectory tracking. Unlike prior studies that focused on~trajectory tracking itself or UAV stabilization control in~simulation, we evaluate the~performance of~a~modified extended IC (eIC) controller compared to~Model Predictive Control (MPC) through both simulated and laboratory experiments with~a~remotely controlled UAV.

The evaluation focuses on~the~computational efficiency and control quality of~real-time UAV trajectory tracking compared to~previous IC applications. The~results demonstrate that the~eIC controller achieves competitive performance compared to~MPC while significantly reducing computational complexity, making it a~promising alternative for~resource-constrained platforms.


\end{abstract}


\section{INTRODUCTION}

In UAV applications, it is important to~ensure that~the~UAV will fly along the~given trajectory. To solve this problem, trajectory tracking methods are used, which are usually implemented on-board in~the~Flight Control Unit or an~additional on-board computing system. These methods are capable of~closed-loop control, which guarantees the~robustness of~the~system, taking into account the~future course of~a~reference trajectory. In the~tracking problem, the~complex dynamics of~a~specific UAV can be considered along with its constraints given by~structural and physical properties.

The most widely used method in~UAV trajectory tracking is the~Model Predictive Control (MPC) \cite{Baca2019,Kamel2017, Nascimento2021}. The~MPC provides a~solution to~the~problem on~the~receding horizon. By~directly incorporating the~prediction into the~control strategy acquisition, the~MPC is able to~consider the~future development of~the~reference trajectory. However, the~consideration of~a~significant part of~the~future trajectory can lead to~a~major increase in~complexity and thus to~much higher computational time demands.

A computationally efficient alternative to~the~MPC can be seen in~Interpolating Control (IC) \cite{nguyen2013constrained, Scialanga2019}. We have already successfully employed IC for~UAV stabilization using explicit IC \cite{boucek-ic-uav}. Unfortunately, the~standard IC was designed only for~control to~the~origin of~the~state space, i.e. stabilization.

Therefore, in~the~paper \cite{boucek-ic-setpoint}, we proposed a~modification of~the~standard IC for~control to~a~constant setpoint control. We have further extended this modification to~include reference trajectory tracking \cite{Boucek2020}. Nonetheless, these modifications have so far been tested only in~a~simulation with a~simple system.

This paper presents an~algorithm based on~IC for~controlling the~UAV along a~given trajectory. The~algorithm constructs a~control strategy that effectively considers future states of~the~UAV. The~resulting algorithm will be tested both in~simulation and in~the~laboratory on~the~Crazyflie UAV\footnote{Crazyflie -- \url{www.bitcraze.io/products/crazyflie-2-1/}}.




\section{\uppercase{Trajectory Tracking for~UAVs}}
\label{sec:track}
The UAV trajectory tracking aims to~follow a~trajectory reflecting the~constraints with high precision. The~constraints take into account the~physical attributes of~the~UAVs and the~restrictions imposed by~the~task, for~example, a~limited rotor speed, a~restricted attitude, speed limits or to~prevent the~UAV from flying into a~restricted area or altitude. 

The trajectory tracking controller is generally implemented directly onboard the~UAV, where a~control code usually runs in~a~loop on~the~processor at a~given frequency, so it is advantageous to~consider the~discrete model of~the~behavior. Considering these attributes, the~trajectory tracking problem is often formulated as an~Optimal Control Problem (OCP) \cite{anderson2007optimal} for~discrete-time linear time-invariant systems with linear constraints and a~quadratic criterion for~evaluation of~the~control quality.

The optimization problem is in~such cases formulated as
\vspace*{-5pt}
\small\begin{align}
	\label{eq:ocpd-crit}
	J\left(\bdx_0,\bdu_0^M\right) &= \left(\bdx_M - \bar{\bdx}_{M}\right)\tp \bdQ \left(\bdx_M - \bar{\bdx}_{M}\right)\nonumber  \\&+ \sum_{k = 0}^{M-1} \left[ \left(\bdx_k - \bar{\bdx}_{k}\right)\tp \bdQ \left(\bdx_k - \bar{\bdx}_{k}\right) +  \bdu_k\tp \bdR \bdu_k\right].\\
	\label{eq:ocpd-sys}
	\text{s.t.}\;\;\;\bdx_{k+1} &= \bdA\bdx_k + \bdB\bdu_k,\, k = 0,1,\ldots,M,	\\
	\label{eq:ocpd-x-const}
	\bdx_k \in \setx,\,\setx &= \left\{\bdx_k\in\real^n:\setxF \bdx_k \leq \setxg\right\},\, k = 0,1,\ldots,M,	\\
	\label{eq:ocpd-u-const}
	\bdu_k \in \setu,\,\setu &= \left\{\bdu_k\in\real^m:\setuF \bdu_k \leq \setug\right\},k = 0,1,\ldots,M,
\end{align}
\normalsize where a~long control horizon $\,M\gg0$ is considered. The~system is controlled along the~given reference trajectory $\bar{\bdx}_{0}^M$, with each reference point along the~trajectory denoted as $\bar\bdx_k$. The~weighting matrices $\bdQ$ and $\bdR$ of~the~quadratic cost function \eqref{eq:ocpd-crit} are known symmetric positive semidefinite and positive definite, respectively. The~quantities $\bdx_k \in \real^{n}$  and $\bdu_k \in \real^{m}$ are a~state and control vector at time instant $k$, respectively.

The optimization constraints \eqref{eq:ocpd-sys}-\eqref{eq:ocpd-u-const} are given by~the~dynamics of~the~UAV, the~considered state space, and feasible control action. State space is constrained by~linear inequality with matrix $\setxF$ and vector $\setxg$; control constraint inequality is defined using matrix $\setuF$ and vector $\setug$.


Many feasible solutions to~the~OCP \eqref{eq:ocpd-crit}-\eqref{eq:ocpd-u-const} are based on~the~employment of~the~standard linear-quadratic regulator (LQR) law \cite{anderson2007optimal}. This control law is optimal for~the~OCP given only by~\eqref{eq:ocpd-crit}-\eqref{eq:ocpd-sys}. The~LQRs used in~this paper for~setpoint control and trajectory tracking are based on~the~description in~\cite{anderson2007optimal} and were presented in~\cite{boucek-ic-setpoint, Boucek2020}.

\subsection{Model Predictive Control}
\label{subsec:mpc}

MPC \cite{borrelli2017predictive} reduces the~complexity of~the~constrained OCP by~solving the~OCP over a~much shorter control horizon and employs a~receding horizon policy, which means that at each time instant only the~control $\bdu_k$, that is given as a~solution to~a~particular OCP at the~time instant $k$, is applied. 

The MPC is the~state-of-the-art method for~trajectory tracking because it uses prediction to~acquire a~control strategy and at the~same time it can take into account given conditions. 
The description of~the~MPC is similar to~the~OCP \eqref{eq:ocpd-crit}-\eqref{eq:ocpd-u-const} with an~adjusted criterion, which is in~each time step given as
\vspace*{-5pt}
\small{
\begin{align}
	\label{eq:mpc-crit}
	J\left(\bdx_{k},\bdu_{k}^{k+N},N\right) = \left(\bdx_N-\bar\bdx_N\right)\tp \bdQ\left(\bdx_N-\bar\bdx_N\right) \nonumber \\ +\sum_{l = k}^{k+N-1} \left[ \left(\bdx_l-\bar\bdx_l\right)\tp \bdQ\left(\bdx_l-\bar\bdx_l\right) + \bdu_l\tp \bdR \bdu_l\right],
\end{align}}
\normalsize
where $N$ is the~length of~the~receding horizon. The~constraints \eqref{eq:ocpd-sys}-\eqref{eq:ocpd-u-const} are the~same as for~OCP.

The solution to~the~MPC can be obtained by~transcribing the~problem to~quadratic programming (QP) \cite{borrelli2017predictive} which can be solved by~a~QP solver. 
The main drawback of~MPC is that the~solution may not be feasible or may be difficult to~obtain.

\subsection{Interpolating Control Based Trajectory Tracking}
\label{subsec:interpolating-control}

IC \cite{nguyen2013constrained,Scialanga2019} is a~promising methodology applicable to~the~OCP. The~major advantage of~IC is the~possibility of~obtaining the~control action by~solving a~very simple linear program (LP). This section describes IC in~terms of~the~IC-based trajectory tracking proposed in~\cite{Boucek2020}.

The IC is based on~the~interpolation between multiple state-feedback gain control laws designed without considering inherent constraints. Using the~invariant set theory \cite{nguyen2013constrained}, IC ensured the~constraints were not violated. 


Initial studies revealed unsatisfactory performance for~standard IC-based UAV trajectory tracking. To address this, we developed a~novel design of~eIC that combines setpoint control and trajectory tracking LQRs. This eIC includes an~additional set with a~setpoint controller, while the~remaining controllers are reflecting the~trajectory.

\subsubsection{Invariant Sets}
\label{sec:inv-sets}

The positively invariant sets used in~IC design are described in~\cite{Boucek2020}. We consider the~system presented in~\Cref{sec:track}. Since the~system \eqref{eq:ocpd-sys} and the~constraints \eqref{eq:ocpd-x-const} and \eqref{eq:ocpd-u-const} are considered linear, the~sets are in~the~form of~polytopes. The~system is controlled by~LQR.





Positively Invariant Set is defined as follows: $\bdvOmega \subseteq \mathcal{O}$ is said to~be positively invariant w.r.t. controlled system in~a~closed loop if and only if $\forall \bdx_k  \in \bdvOmega$. This implies that once the~state $\forall \bdx_k$ reaches $\bdvOmega$, it will remain within $\bdvOmega$ while satisfying the~state and control constraints.

\subsubsection{State Decomposition}
\label{sec:state-decomp}

The interpolation employs the~principle of~state decomposition, which can be denoted as
\vspace*{-5pt}
\begin{equation}
\label{eq:ic-decomp-state}
\bdx = c \bdx^l + \left(1-c\right)\bdx^h,
\end{equation}
where $\bdx$ is the~state vector, $c$ is the~interpolating coefficient, $c\in \left< 0,1 \right>$, and $\bdx^h$ and $\bdx^l$ is the~state vector for~a~high-gain and a~low-gain controller, respectively. In the~case of~trajectory tracking, the~decomposed state from \Cref{eq:ic-decomp-state} is reflected in~the~IC law as
\vspace*{-5pt}
\begin{equation}
\label{eq:ic-decomp-ctrl}
\medmath{
\bdu (\bdx_k, \bar{\bdx}_{k}^{k+N}) = c_k \bdu^l \left( \bdx^l_k, c_k \bar{\bdx}_{k}^{k+N}\right) + \left(1-c_k\right) \bdu^h \left( \bdx^h_k, \left(1-c_k \right) \bar{\bdx}_{k}^{k+N}\right),}
\end{equation}
where $\bdu^h(\bdx^h_k, \bar{\bdx}_{k}^{k+N})$ is the~high-gain control law for~$\bdx^h_k$ and $\bdu^l(\bdx^l_k, \bar{\bdx}_{k}^{k+N})$ is the~low-gain control law for~$\bdx^l_k$. The~setpoint control law is obtained by~substituting $\bar{\bdx}_{k}^{k+N}$ with $\bar{\bdx}_{k}$.



While both controllers could be used by~simply switching between them based on~the~system's current state, it would introduce discontinuity in~the~control. Moreover, in~the~region $\bdvOmega^l\setminus\bdvOmega^h$, it would lead to~slower convergence of~the~system state towards the~desired region of~the~state space.

\subsubsection{Interpolating Control}
\label{sec:ic-implicit}

As has been said, the~standard IC depends on~finding the~optimal interpolation coefficient $c^*$, which can be acquired by~minimizing the~criterion in~simple LP, which was described for~setpoint control in~\cite{boucek-ic-setpoint} and for~trajectory tracking in~\cite{Boucek2020}. The~LP adjusts to~setpoint control or trajectory tracking by~shifting the~center of~set $\bdvOmega^h$ of~the~high-gain controller $\bdu^h(\bdx^h_k, \bar{\bdx}_{k}^{k+N})$ from the~origin coordinates to~the~setpoint or current point of~reference trajectory coordinates.



\subsubsection{Extended Interpolating Control}

In case the~set $\bdvOmega^l\setminus\bdvOmega^h$ is large, the~performance of~the~IC can be degraded. The~performance can be usually improved by~adding an~intermediate invariant set $\bdvOmega^l \subset \bdvOmega^m \subset \bdvOmega^h$, where another intermediate state-feedback controller is defined. This extended version of~IC is called eIC for~clarity. In case $\bdu^h$ is LQR, $\bdu^m$ can be obtained for~example with an~increase in~the~weight $\bdR$. The~set $\bdvOmega^m$ is calculated as $\bdvOmega^h$. 


The eIC modification for~trajectory tracking presents challenges, as discussed in~\cite{Boucek2020}. To ensure that $\bar{\bdx}_{k}^{k+N} \in \bdvOmega^m$ when controlling within the~region $\bdvOmega^m\setminus\bdvOmega^h$, adjustments must be made to~the~reference trajectory. These adjustments become even more complex because the~center of~set $\bdvOmega^m$ is also shifted to~the~coordinates of~the~current reference trajectory point. 

In eIC, there are two distinct LPs. If $\bdx \in \bdvOmega^l\setminus \bdvOmega^m$, the~interpolation is done between $\bdu^l$ and $\bdu^m$ and the~IC control law takes the~form of
\vspace*{-5pt}
\begin{equation}
\medmath{
	\label{eq:eic1-decomp-ctrl}
	\bdu (\bdx_k, \bar{\bdx}_{k}^N) = c_k \bdu^l \left( \bdx^{l}_k, c_k \bar{\bdx}_{k}^N \right) + \left(1-c\right) \bdu^m \left( \bdx^m, \left(1-c_k\right) \bar{\bdx}_{k} \right).
 }
\end{equation}
If the~$\bdx\in\bdvOmega^m$, the~interpolation is performed for~both $\bdu^m$ and $\bdu^h$, which results in~control law
\vspace*{-5pt}
\begin{equation}
\medmath{
	\label{eq:eic2-decomp-ctrl}
	\bdu (\bdx_k, \bar{\bdx}_{k}^N) = c_k \bdu^m \left( \bdx^m_k, c_k\bar{\bdx}_{k} \right) + \left(1-c_k\right) \bdu^h\left( \bdx^h_k, \left(1-c_k\right) \bar{\bdx}_{k}^N\right).
 }
\end{equation}

\section{\uppercase{Planar UAV Model and Control Design}}

Both MPC and IC are model-based methods. Their design requires knowledge of~the~behavior model of~the~controlled system. Therefore, this section outlines the~behavioral model of~UAVs.  Additionally, it outlines constraints based on~the~UAV's characteristics and flight space. Finally, it describes the~parameters of~each UAV controller.

\subsection{Planar UAV Model}

\begin{figure}[thpb]
    \centering
	   \input{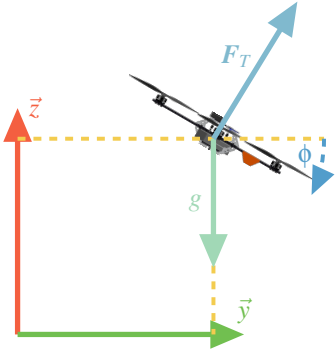}
	\caption{Planar UAV in~the~local frame}
	\label{fig:planar}
   \end{figure}

For easier analysis and better insight, a~planar UAV mo\-de\-l (see \Cref{fig:planar}) will be employed, which exhibits similar behavior but is reduced both in~the~number of~state variables and in~the~complexity of~the~equations of~motion.  The~dy\-na\-mics along $\vecy$ and $\vecz$-axis with attitude $\phi$ as rotation around axis $\vecx$ are considered.

The nonlinear dynamics of~the~planar UAV is described by~the~following equations
\vspace*{-2pt}
\begin{align}
	\label{eq:uav-planar-nonlin}
 \medmath{
	\ddot{\y}(t) =  -\dfrac{F_T(t)}{m}\sin(\phi(t)),  \,
	\ddot{\z}(t) =  -g + \dfrac{F_T}{m}\cos(\phi(t)), \,
	\ddot{\phi}(t) =  \dfrac{\torquecollectivex(t)}{J_x},}
\end{align}
where $F_T$ is the~collective thrust in~[N], $m$ is the~mass of~the~UAV in~[kg], $\torquecollectivex$ is the~collective torque in~$\left[\mathrm{N\cdot m}\right]$ generated by~rotors around $\vecx$-axis and $J_x$ is the~moment of~inertia around $\vecx$ in~$\left[\mathrm{kg\cdot m^2}\right]$.

The attitude control is handled by~the~UAV's autopilot, the~proposed controller can control the~translation in~$\vecy$ and $\vecz$ via desired acceleration, which is recalculated based on~linearized equations \eqref{eq:uav-planar-nonlin} around hover state to~attitude control reference as
\vspace*{-5pt}
\begin{align}
	\label{eq:uav-planar-transform}
	\bar{\phi}(t)=-\dfrac{\ddot{\bar y}(t)}{g},\,F_{T}(t)=m\left(\ddot{\bar z}(t) + g\right),
\end{align}
where $\bar\phi$ and $F_{T}$ are desired angle and collective thrust, respectively. Moreover, as a~side effect, the~controllers are independent of~the~UAV's parameters $m$ and $J_x$, as they are only compensated in~the~calculations for~the~attitude and thrust control setpoint. 


\subsection{Model Parameters}

\begin{figure}[thpb]
	\centering
    \parbox{3in}{
    \begin{subfigure}[t]{1.45in}
        \centering
        \includegraphics[width = \textwidth]{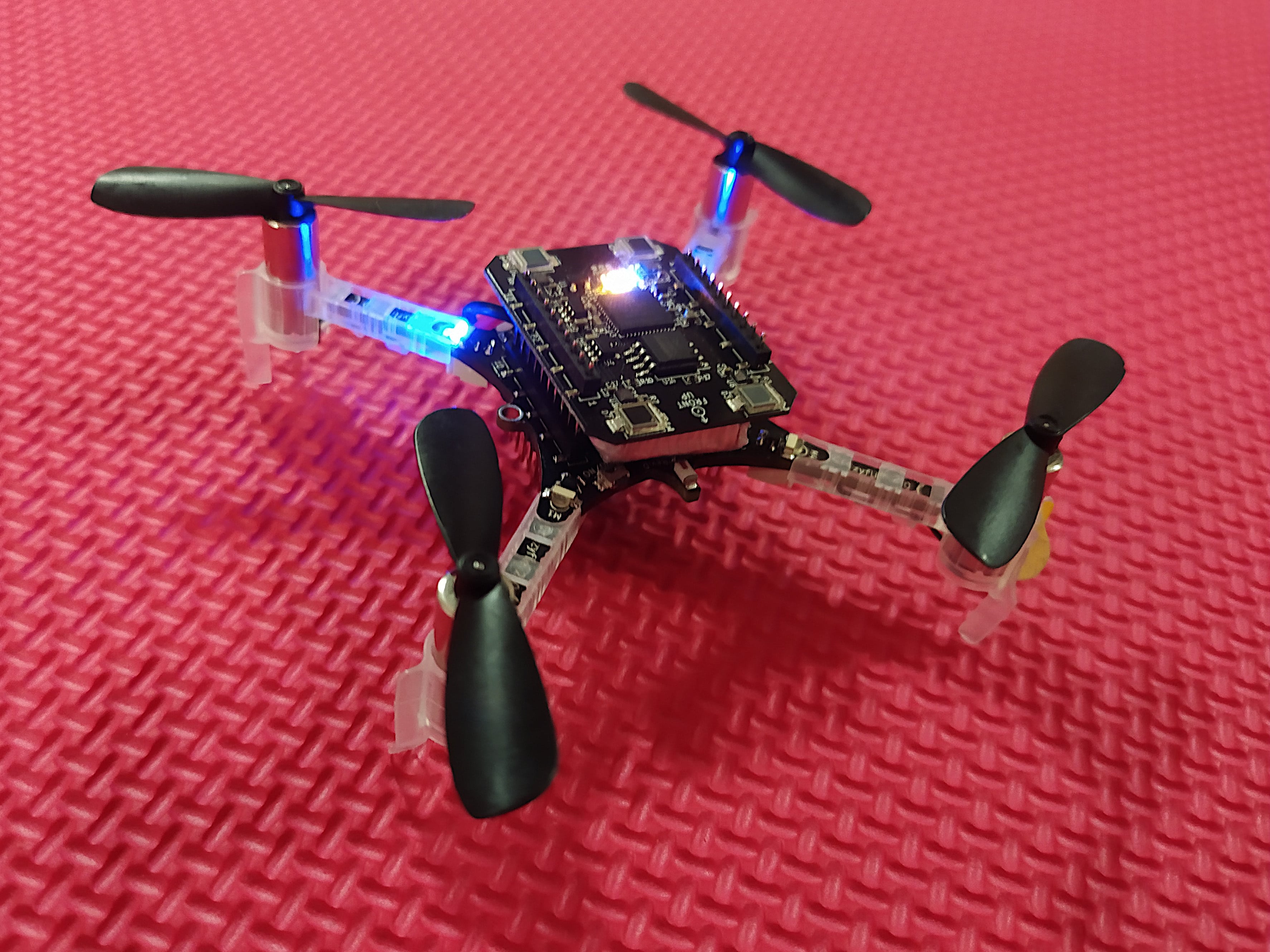}
	\caption{Crazyflie UAV}
	\label{fig:crazyflie-top}
    \end{subfigure}%
    \hfill
    \begin{subfigure}[t]{1.45in}
        \centering
	   \includegraphics[width = \textwidth]{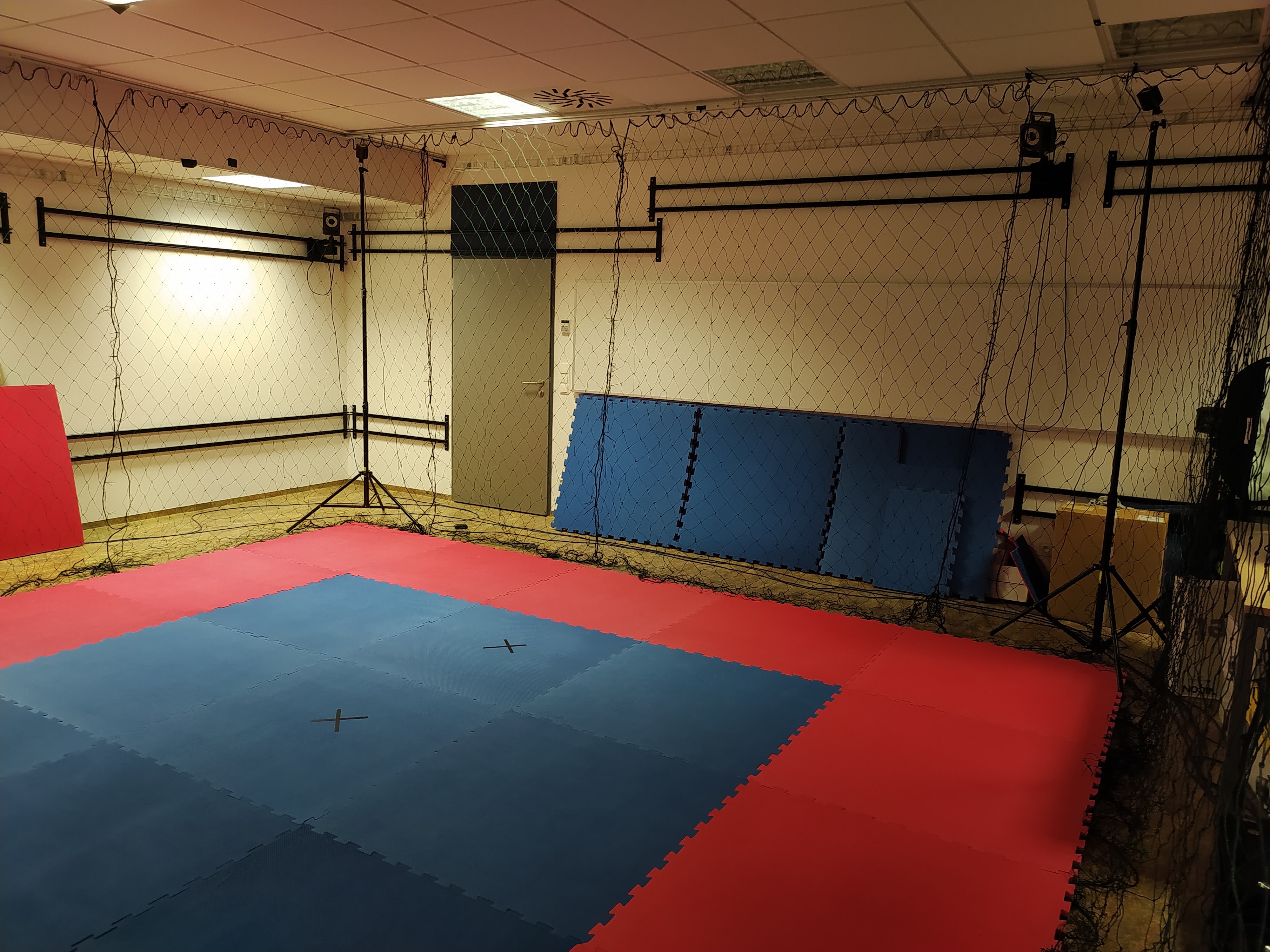}
    \caption{Flight arena}
	\label{fig:lab}
    \end{subfigure}}
    \caption{Laboratory Experiment with Crazyflie UAV}
\end{figure}

The model parameters were selected to~align with a~real UAV, specifically, the~Crazyflie 2.0 developed by~Bitcraze (see \Cref{fig:crazyflie-top}). Similarly, the~state and control constraints were set using both experimentally measured and manufacturer-provided values. Regarding the~model's parameters, we set the~UAV's mass to~$m=0.03\, \mathrm{kg}$ and the~moment of~inertia to~$J_x =
2.3951\cdot10^{-5}\, \mathrm{kg\cdot m^2}$. State and control constraints were defined as
\vspace{-5pt}
\[
\begin{aligned}
	-2 \leq   & y \leq 2, && -1.25 \leq  z \leq 1.25, && \mathrm{[m]} ,\\
	-5 \leq &\dot{y} \leq 5,  && \phantom{0}\phantom{0}-5 \leq \dot{z} \leq 5, && \mathrm{[m/s]},
\end{aligned}
\]
where the~position constraints align with the~laboratory parameters (see \Cref{fig:lab}), and the~velocity constraints are set to~allow high-performance maneuvers. The~lower bound for~the~position in~the~$\vecz$-axis is negative to~enable stabilization control, while $\vecz$ is transformed to~have system origin at a~height of~1.25 m.

\subsection{Controllers Design}

The weights of~the~quadratic criterion used in~designing the~controllers were appropriately selected based on~the~UAV's behavior and surrounding environment as
\vspace*{-5pt}
\[
\begin{aligned}
	\bdQ_y^h =&  \begin{bmatrix} 0.16 & 0 \\ 0 & 0.04 \end{bmatrix}, \,
	\bdQ_z^h =  \begin{bmatrix} 0.64 & 0 \\ 0 & 0.04 \end{bmatrix}, \\
    \bdR_y^h =& \frac{1}{2} \cdot 0.15 \cdot \frac{4}{0.03} \cdot \sin{30^\circ} = 5,\, \bdR_z^h = 5^{-2} = 0.04,
\end{aligned}
\]
where $\bdQ_y^h$, $\bdR_y^h$, $\bdQ_z^h$, and $\bdR_z^h$ are the~weights utilized in~MPC, IC, and eIC for~position control in~$\vecy$ and $\vecz$-axis. 

For trajectory tracking, IC employs the~LQR $\bdu^l$ with the~following weight configuration:
\vspace*{-0.5em}
\[
\begin{aligned}
	\bdQ_y^l =& \begin{bmatrix} 0.25 & 0 \\ 0 & 0.04 \end{bmatrix}, \,
	\bdQ_z^l = \begin{bmatrix} 0.64 & 0 \\ 0 & 0.04 \end{bmatrix}, \\ 
    \bdR_y^l =& 10\cdot \bdR_y^h = 50,\, \bdR_z^l =10\cdot \bdR_z^h = 0.4.
\end{aligned}
\]
The setpoint LQR $\bdu^m$ uses the~following weighting matrices
\begin{equation}
	\bdQ_y^m = \bdQ_y^l,\,\bdQ_z^m = \bdQ_y^l,\, \bdR_y^m =  \bdR_y^h,\, \bdR_z^m = \bdR_z^h.
\end{equation}

The predictive horizon was set to~8s as in~\cite{Baca2019}, resulting in~$N=800$ steps for~the~$T_s=0.01$s discrete model. However, such a~long horizon is not feasible for~real-time MPC implementation. To address the~issue, we employed the~move blocking technique \cite{Cagienard2007}, which fixes the~control variable values  $\bdu^{k+N}_k$  for~multiple time steps, reducing the~QP's degrees of~freedom. To further reduce complexity, the~prediction for~$T_s=0.01$s is performed only in~the~first step and afterward, the~model with $T_s = 0.2$s is considered. The~complexity of~the~problem was reduced by~95\%. This version of~the~MPC will be referred to~hereafter as MPCMB. 

A basic discrete-time PID controller is implemented in~Python with experimentally obtained parameters $K_p = 0.3$, $K_d=0.003$, and $K_i = 0.0001$, and with a~period $T_{s\_att}=0.001$s. The~PID simulates the~attitude control behavior in~a~scenario with the~planar UAV model, which would otherwise be done by~the~autopilot. 

\section{\uppercase{Experimental Setup}}

The experiments are conducted using both a~planar model simulation and a~3D simulation environment Gym-PyBullet-Drones \cite{panerati2021learning}, and with the~Crazyflie UAV wirelessly controlled in~the~laboratory utilizing the~cflib library\footnote{CFLib -- \url{https://github.com/bitcraze/crazyflie-lib-python}}. The~MPC has been excluded from real-world and Gym-Pybullet-Drones tests due to~its high computational complexity.


Simulations are executed on~a~desktop PC with an~Intel Core i9-9900  and 64GB  DDR4 RAM,  running Ubuntu 22 and Python 3.8. The~LPs and QPs were modeled using CVXPY and solved using GUROBI \cite{gurobi}. In the~laboratory experiment, a~laptop with Intel Core i7-8550  and 16GB DDR3 RAM is used with the~same software setup.

Tracking was performed for~two types of~reference trajectories: the~Ellipse and the~Lemniscate of~Gerono. As both trajectories delivered similar results, only the~Lemniscate is presented. 

\subsection{Simulations}

The planar model is implemented in~Python based on~the~nonlinear dynamics described in~\eqref{eq:uav-planar-nonlin}. The~reference signal for~lemniscate was generated for~$\vecx$ and $\vecy$ axes. Other reference states are equal to~zero vectors, however, since the~reference signal is time-parameterized, it is, therefore, the~trajectory.

The response is tested for~angular frequencies $\omega_s = 0.4$ and  $\omega_s = 0.6$. For $\omega_s = 0.6$, the~reference is referred to~as a~high-frequency lemniscate trajectory. The~simulation begins with the~initial condition at the~state space origin, with a~nonzero reference requiring initial convergence.

The Gym-PyBullet-Drones \cite{panerati2021learning} is a~simulation environment for~single and multi-agent reinforcement learning with na\-no quad\-cop\-ters. To test controllers under more practical circumstances, this environment features the~simulation of~Crazyflie UAV. The~lemniscate trajectory is set in~the~$\vecx$ and $\vecy$ axes, while the~$\vecz$-axis is set to~a~faster-oscillating signal 
to provide a~more complex test of~the~$\vecz$-controller.

Since this is a~3D simulation, the~$\vecy$-axis controller is also used for~controlling the~$\vecx$-axis. The~output of~the~$\vecx$ and $\vecy$ controllers must be rotated around the~$\vecz$-axis in~order reflect the~UAV's yaw angle $\psi$.

\subsection{Laboratory Experiment with Crazyflie UAV}

The laboratory experiments use Crazyflie 2.0 UAVs equipped with the~Lighthouse positioning deck\footnote{LH deck -- \url{www.bitcraze.io/lighthouse-positioning-deck}}, which enables UAV's self-localization via the~HTC SteamVR Base Station 2.0 with high accuracy. 

MPCMB exhibits excessively long computation times during experiments, exceeding the~expected $T_s=0.01$s. The~UAV is unable to~stabilize in~the~$\vecz$-axis and keeps oscillating, resulting in~a~crash. To address this issue, two types of~experiments are conducted: First, MPCMB generates roll and pitch angle and the~$\bar z$ position is directly set using \texttt{send\_zdistance\_setpoint} function. Secondly, the~thrust is controlled by~ICs using the~function \texttt{send\_setpoint}. For ICs, parallel computing is also tested using Pool from the~multiprocessing library. The~trajectory was adjusted for~safety purposes and was slightly smaller. 


\section{RESULTS AND DISCUSSION}

The controllers are evaluated for~tracking a~lemniscate reference trajectory in~terms of~control quality and computational complexity. Control quality is assessed based on~the~quadratic criterion, the~integral square error (ISE), and the~energy required to~control the~UAV.

The quadratic criterion value in~\Cref{eq:ocpd-crit} is weighted by~the~inverse of~$T_s$. In contrast, the~ISE evaluates the~accuracy of~the~reference position tracking. For example, for~position $x$ the~ISE is calculated as
\vspace*{-5pt}
\begin{equation}
	\label{eq:ic-ise}
	ISE = \dfrac{1}{T_s}\sum_{k=0}{e_{k}^2},
\end{equation}
where $e_{k} = \left(\bar x_{k}-x_{k}\right)$. The~presented ISE value is the~sum of~the~ISE from all axes. The~energy is directly dependent on~the~controller output, which is by~design given as the~desired acceleration. To illustrate, for~the~$\vecz$-axis controller, energy is assessed using the~following equation
\vspace*{-5pt}
\begin{equation}
	\label{eq:ic-energy}
	E = \dfrac{1}{T_s}\sum_{k=0}{\ddot\bar z_k^2}.
\end{equation}

In laboratory experiments, the~criteria cannot be weighted solely by~$T_s$ due to~asynchronous communication and delays resulting from the~control action calculation. Thus, the~difference between time instants $t_k-t_{k-1}$ is used to~weight the~individual elements. The evaluation of~controller time demands is based on~total, average, and maximum computation time. For laboratory experiments, the~number of~calculated control actions is also included.


\subsection{Simulations}

When $\omega_s=0.4$ s, the~controllers exhibit similar behavior. The~difference is observed only at the~beginning, where ICs and MPCs have quite different paths. The~interpolating coefficients are zero except at the~very beginning of~the~simulation in~the~$\vecy$-axis. Thus, the~ICs achieved optimal behavior for~the~most of~time. It \Cref{tab:comparison-crit-sim-figure8}, we can see the~interpolation-based controllers are worse in~optimality criterion because they consumed more energy. However, they followed the~trajectory more closely.

\begin{figure}[thpb]
	\centering
	\parbox{3in}{\includegraphics[width = 0.9\linewidth]{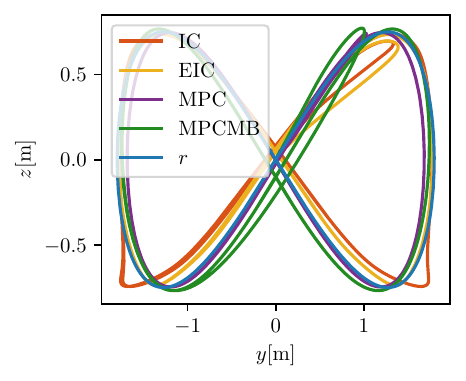}}
	\caption{Path from tracking the~high-frequency trajectory with the~planar model}
	\label{fig:uav-track-traj-fig8-faster}
\end{figure}

Very different results were obtained for~the~high-frequency trajectory with $\omega_s=0.6$ s. According to~\Cref{fig:uav-track-traj-fig8-faster}, the~ICs again tried to~follow the~reference faster. The~IC deviates significantly at two points, possibly due to~the~interaction between controllers, which may cause this issue. As the~$\vecy^\localframe$ controller increases the~$\phi$ angle for~greater acceleration, it can lead to~a~deflection of~the~thrust controlled by~the~$\vecz^\localframe$ controller. According to~the~optimality criterion, \Cref{tab:comparison-crit-sim-figure8} demonstrates that the~MPC produced the~most favorable outcomes. However, the~eIC again followed the~trajectory closest. In contrast, the~IC achieved the~worst results by~all criteria.

\begin{table}
	\centering
\caption{Evaluation of control quality with the planar UAV model (2D) and Gym-PyBullet-Drones (3D) simulation}
\label{tab:comparison-crit-sim-figure8}
\resizebox{\linewidth}{!}{%
\begin{tabular}{lrrrrrr}
    \hline
\textbf{2D $\omega_s=0.4$}& $J$  & $\%$      & $ISE$  & $\%$       & $E$ & $\%$ \\
\hline
MPC & 7.07 & - & 1.76 & - & 21.20 & - \\
MPCMB & 7.74 & +9.48 & 2.19 & +24.43 & 16.80 & -20.75 \\
eIC & 8.23 & +16.41 & 1.73 & -1.70 & 27.30 & +28.77 \\
IC & 8.68 & +22.77 & 1.50 & -14.77 & 33.80 & +59.43 \\
\hline
\textbf{2D $\omega_s=0.6$}& $J$              & $\%$      & $ISE$  & $\%$       & $E$                 & $\%$ \\
\hline
MPC & 13.60 & - & 2.02 & - & 61.20 & - \\
MPCMB & 15.70 & +15.44 & 2.64 & +30.69 & 58.30 & -4.74 \\
eIC & 16.40 & +20.59 & 1.73 & -14.36 & 79.30 & +29.58 \\
IC & 23.50 & +72.79 & 3.86 & +91.09 & 98.00 & +60.13 \\
\hline
\textbf{3D $\omega_s=0.6$}& $J$              & $\%$      & $ISE$  & $\%$       & $E$                 & $\%$ \\
\hline
MPCMB & 52.8 & - & 6.40 & - & 2.79 & - \\
eIC & 73.8 & +39.77 & 2.71 & -57.66 & 3.77 & +35.13 \\
IC & 192 & +263.64 & 5.20 & -18.75 & 4.68 & +67.74 \\
\hline
\end{tabular}}
\end{table}

In the~3D simulation environment, the~results for~$\omega_s = 0.4$ are comparable to~those of~the~planar model, only the~case of~the~high-frequency lemniscate will be presented. The~MPCMB achieved the~best optimality criterion value, as seen in~\Cref{tab:comparison-crit-sim-figure8}. The~IC was too aggressive, consuming excessive energy, and its optimality criterion value was 264\% higher than the~MPCMB. In contrast, the~eIC performed 40\% worse than the~MPCMB in~terms of~the~optimality criterion but had precise trajectory tracking. The~eIC attained substantially lower ISE than the~MPCMB while maintaining reasonable energy use.





\subsection{Laboratory Experiments with Crazyflie UAV}

\begin{figure}
	\centering
	\parbox{3in}{\includegraphics[width=0.9\linewidth]{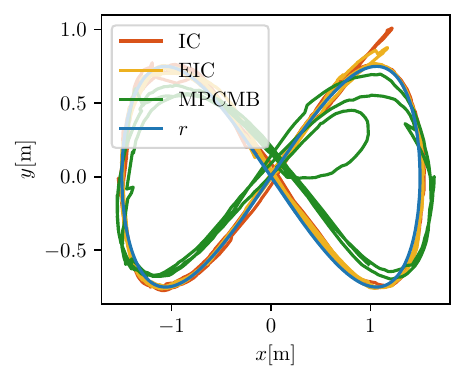}}
	\caption{Path from tracking the~high-frequency trajectory with Crazyflie UAV in~laboratory}
	\label{fig:uav-track-lab-traj-fig8}
\end{figure}

Since the~controllers only handle Crazyflie's motion in~the~$\vecx^\localframe$ and $\vecy^\localframe$ axes, the~presented plots show the~controllers' performance exclusively in~those axes. During lab testing, the~ICs promptly followed the~reference (\Cref{fig:uav-track-lab-traj-fig8}). In contrast, the~MPCMB maneuvered the~path at a~slower pace, resulting in~a~smaller path. 
The MPCMB attained the~lowest cost, while the~ICs had 10--15\% higher values (\Cref{tab:comparison-crit-lab-figure8}). Nevertheless, both ICs tracked the~trajectory more accurately based on~the~ISE, but at significantly higher energy consumption.

\begin{table}
    \centering
\caption{The control quality in the laboratory experiment}
\label{tab:comparison-crit-lab-figure8}
\begin{tabular}{lcccccc}
\hline
\textbf{2D} & $J$  & $\%$      & $ISE$  & $\%$       & $E$ & $\%$ \\
\hline
MPCMB & 9.67 & - & 3.85 & - & 0.42 & - \\
eIC & 10.70 & +10.65 & 1.42 & -63.12 & 1.09 & +161.39 \\
IC & 11.10 & +14.79 & 1.39 & -63.90 & 1.17 & +180.58 \\
\hline
\textbf{3D} & $J$  & $\%$      & $ISE$  & $\%$       & $E$ & $\%$ \\
\hline
eIC par & 20.2 & - & 1.56 & - & 1.62 & - \\
IC par & 21.1 & +4.46 & 1.61 & +3.21 & 1.65 & +1.85 \\
eIC & 19.9 & -1.49 & 1.57 & +0.64 & 1.68 & +3.70 \\
IC & 24.7 & +22.28 & 1.73 & +10.90 & 2.09 & +29.01 \\
\hline
\end{tabular}
\end{table}

\begin{figure}
    \centering
    \parbox{3in}{\includegraphics[width=\linewidth]{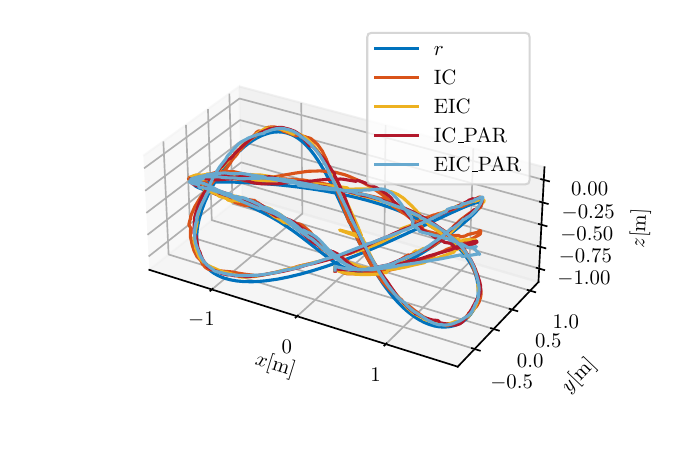}}
    \label{fig:uav-track-lab-traj-3d-fig8-thrust+par}
    \caption{Path from 3D tracking the~high-frequency trajectory with Crazyflie UAV in~laboratory}
	\label{fig:uav-track-lab-traj-fig8-thrust+par}
\end{figure}

For the~3D control, the~captured UAV path is very similar for~all ICs (\Cref{fig:uav-track-lab-traj-fig8-thrust+par}). There are minor differences between the~standard and parallelized versions of~the~controllers. 
Furthermore, we can see a~significant change in~one point in~case of~standard eIC. Such a~sudden change can be explained by~a~poor estimate of~the~UAV's position. However, all ICs successfully track the~reference trajectory.

The performance evaluation shown in~\Cref{tab:comparison-crit-lab-figure8} indicates that both eICs yield comparable results. That said, the~standard eIC outperforms the~parallel eIC in~terms of~the~optimality criterion, despite consuming more energy and deviating slightly according to~ISE. The~values of~the~weighting matrices and the~small differences in~the~behavior of~both controllers cause this paradox. Interestingly, the~parallel IC is only 4.5\% worse than the~parallel eIC, which is even more intriguing since the~standard IC is 22.3\% worse.

\subsection{Computational Demands}
The computational demands are consistent for~all scenarios. Therefore, only the~results for~the~high-frequency lemniscate trajectory are presented.

\begin{figure}
	\centering
	\parbox{3in}{\includegraphics[width=0.9\linewidth]{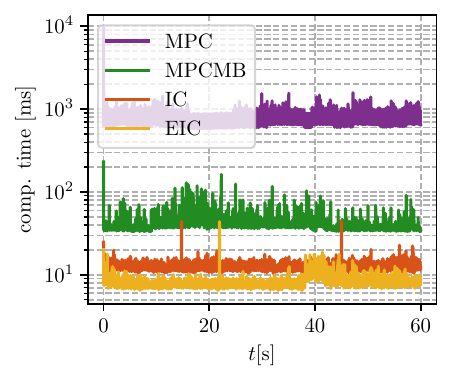}}
	\caption{Computational time for~tracking the~fast lemniscate trajectory with a~planar model of~the~UAV}
	\label{fig:uav-track-2nonlin-time-fast-fig8}
\end{figure}

The logarithmic plot in~\Cref{fig:uav-track-2nonlin-time-fast-fig8} indicates the~MPC took the~longest computation time, followed by~MPCMB. IC and eIC have comparable demands with eIC faster the~most of~time, likely due to~the~shape of~the~$\bdvOmega^m$ set and LP. \Cref{tab:comparison-time-2d-figure8-fast} confirms the~order of~magnitude differences, also for~3D simulation. The~findings demonstrate that IC is a~highly time-efficient substitute for~MPC. For MPCs, it is evident that the~initial computational time is significantly longer than subsequent times. This may be due to~the~solver utilizing the~previous step's results in~the~following step.

\begin{table}
	\centering
\caption{The time demands with the planar UAV model (2D) and Gym-PyBullet-Drones (3D) simulation}
\label{tab:comparison-time-2d-figure8-fast}
\begin{tabular}{lcccc}
    \hline
\textbf{2D}& $t$ [s] & $\%$    & $t_{\max}$ [ms] & $\%$    \\
\hline
MPC & 4180 & - & 9507 & - \\
MPCMB & 247 & -94.09 & 230 & -97.58 \\
eIC & 49 & -98.83 & 44 & -99.54 \\
IC & 75 & -98.21 & 45 & -99.53 \\    
\hline
\textbf{3D}& $t$ [s] & $\%$    & $t_{\max}$ [ms] & $\%$    \\
\hline
MPCMB & 335 & - & 374 & - \\
eIC & 72 & -78.51 & 68 & -81.82 \\
IC & 109 & -67.46 & 51 & -86.36 \\
\hline
\end{tabular}
\end{table}


In the~lab tests, a~noticeable difference is observed between the~ICs (see \Cref{tab:comparison-time-lab-figure8}). It is important to~note that the~previous tests were conducted on~a~desktop computer, while the~lab test was conducted on~a~laptop. Parallelized IC demonstrates lower maximum times, whereas parallel eIC has slightly higher averages. Nonetheless, parallel IC exhibited improved performance over standard IC based on~all metrics. Overall, the~standard eIC remained the~most efficient.


\begin{table}
    \centering
    \caption{The time demands in the laboratory experiment}
    \label{tab:comparison-time-lab-figure8}
\resizebox{\linewidth}{!}{%
    \begin{tabular}{lcccccccc}
    \hline
    \textbf{2D} & $t$ [s] & $\%$ & $N_s$ & $\%$  & $\bar{t}$ [ms] & $\%$  & $t_{\max}$ [ms] & $\%$ \\
    \hline
    MPCMB & 30 & - & 582 & - & 51 & - & 432 & - \\
    eIC & 26 & -13.33 & 2658 & +356.70 & 10 & -80.39 & 118 & -72.69 \\
    IC & 30 & +0.00 & 1407 & +141.75 & 21 & -58.82 & 120 & -72.22 \\
    \hline 
    \textbf{3D}& $t$ [s] & $\%$ & $N_s$ & $\%$  & $\bar{t}$ [ms] & $\%$  & $t_{\max}$ [ms] & $\%$ \\
    \hline
    eIC par & 29  & - & 1648 & - & 18 & - & 73 & - \\
    IC par & 30  & +3.45 & 1153 & -30.04 & 26 & +44.44& 67 & -8.22 \\
    eIC & 30 & +3.45 & 2106 & +27.79 & 14 & -22.22 & 106 & +45.21 \\
    IC & 30 & +3.45 & 993 & -39.75 & 30 & +66.67 & 125 & +71.23 \\
    \hline
\end{tabular}}
\end{table}






   
\section{CONCLUSION AND FUTURE WORK}

The UAV trajectory tracking controllers underwent both simulated and laboratory testing. The~evaluation compared controller performance in~both control quality and computational complexity. Generally, MPCs showed superior performance, especially in~terms of~optimality. Even so, ICs can provide comparable performance and serve as a~viable alternative to~ MPCs due to~their computational efficiency, especially on~devices with limited computing power.

Our research suggests that the~eIC is a~promising controller for~trajectory tracking, displaying faster computation times and enhanced accuracy in~comparison to~IC. 
Additionally, parallel versions of~ICs have shown further reductions in~computation time.
Integrating the~IC directly into the~autopilot of~the~Crazyflie UAV in~the~future could potentially advance trajectory tracking. Although, careful consideration of~the~platform's limitations would be necessary.

\bibliographystyle{IEEEtran}
\bibliography{IEEEabrv,bib-uav-track}

\end{document}